# Comparing BB84 and Authentication-Aided Kak's Three-Stage Quantum Protocol

Partha Basuchowdhuri

**Abstract:** This paper compares the popular quantum key distribution (QKD) protocol BB84 with the more recent Kak's three-stage protocol and the latter is shown to be more secure. A theoretical representation of an authentication-aided version of Kak's three-stage protocol is provided that makes it possible to deal with man-in-the-middle attack.

## 1. Introduction

Although practical quantum computing may be years away, quantum cryptography is already a reality and products are out in the market pioneered by firms like Quantique, MagiQ Technologies, SmartQuantum and there are research groups working on it at corporations like HP, IBM, and Toshiba [2].

Security in software or online applications is generally provided by classical encryption algorithms like Diffie-Hellman, El-Gamal, RSA, and ECC [11], but all of these are either breakable or subject to cryptanalysis if sufficient technological resources were available. In quantum cryptography, on the other hand, it can be proved that Eve (the eavesdropper) cannot access information without Alice (sender) and Bob's (recipient) knowledge.

In 1984, Bennett and Brassard published their BB84 protocol [6], which is now the most popular QKD method. BB84 and its variants use quantum bits in one pass and this is followed by two additional passes of classical data transmission (that are potential weak links). Kak's protocol [1], on the other hand, uses quantum information transmission in all its steps to ensure that there is no weak link in the process. The weakness of the classical data links of the BB84 is apparent from the fact that single photons are not easy to produce [2], and the duplicate photons can be used by the eavesdropper to reconstruct the key.

Corresponding e-mail: pbasuc1@lsu.edu (Department of Electrical and Computer Engineering, Louisiana State University, LA, USA)

2. **Viewing BB84 as a three-stage protocol**

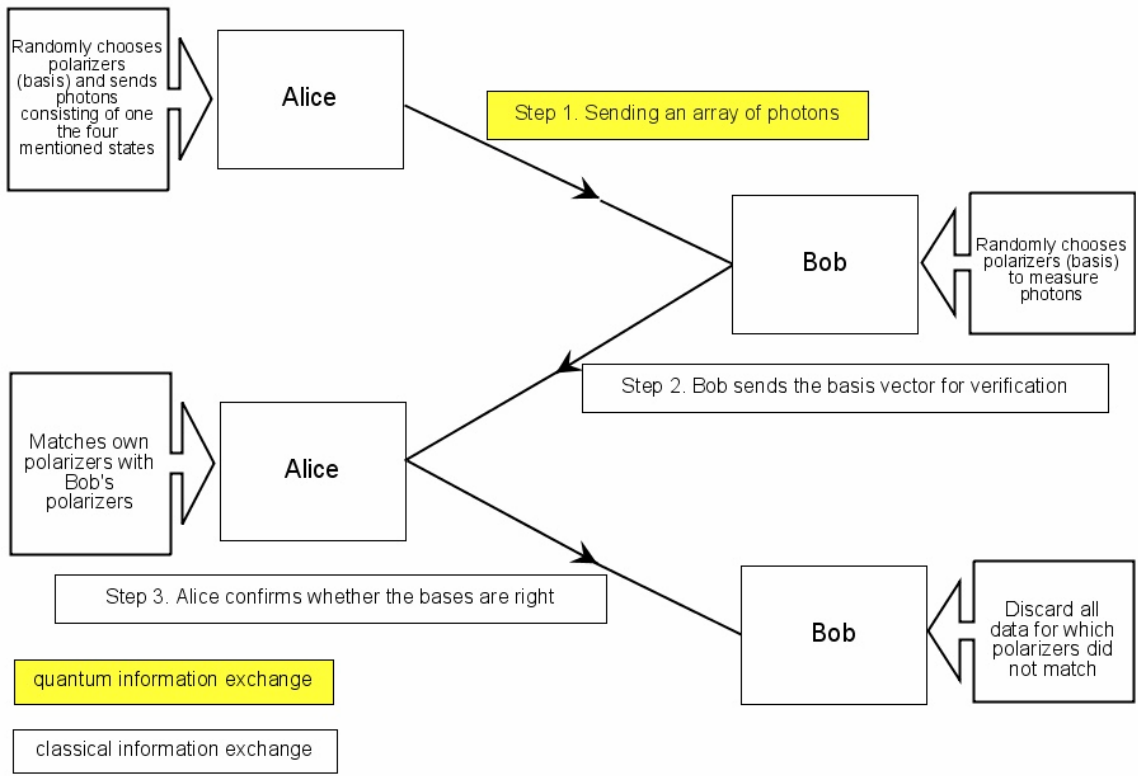

**Fig 1. BB84, viewed as a three-stage protocol**

Say, Alice and Bob each has two polarizers, with 0/90 degrees ( $+$ ) for which $|$ is 1 and $—$ is considered as 0 and with 45/135 degrees ( $\times$ ) for which $\setminus$ is 1 and $/$ is 0.

BB84 protocol can be viewed as a three-stage protocol as said below:

*Step 1:* Alice randomly chooses polarizers to generate photons and sends them to Bob.
Say polarizers chosen by Alice are:   $+$ $\times$ $\times$ $+$ $+$ $\times$ $+$
Say photons sent by Alice are:   $—$ $\setminus$ $/$ $|$ $|$ $/$ $—$

*Step 2:* Bob receives those photons with randomly chosen polarizers.
Photons sent by Alice are:   $—$ $\setminus$ $/$ $|$ $|$ $/$ $—$

Say polarizers chosen by Bob are:   $+$ $+$ $\times$ $+$ $\times$ $\times$ $+$
Bob's resulting measurement is:   $—$ $|$ $/$ $|$ $\setminus$ $/$ $—$

***Step 3:*** Alice and Bob matches their bases and discard the data for un-matched polarizers. So final measurement should be:　　　—　　╱　│　　╱　—
So the resultant bit representation will be: 0-01-00

The main problem for BB84 is that generating single photon is not easy [2]. In most industrial applications weak laser beam is used to send the bits needed for the key. Usually the laser pulses generate two or more photons which remain in the same quantum state. In a beam-splitting attack Eve could split the beam and use the spilt photon to detect the bit and could only let pass the multiple photon beams to Bob. In this way Eve can eventually determine all the key bits and also remain undetected.

## 3. The basic idea of quantum key distribution

Say an eavesdropper Eve is trying to intercept the message being transmitted between Alice and Bob. Quantum key distribution is effective because of the no-cloning theorem [6]. If Eve tries to differentiate between two non-orthogonal states, it is not possible to achieve information gain without collapsing the state of at least one of them.

This is clear from considering $|\psi\rangle$ and $|\varphi\rangle$ to be the non-orthogonal quantum states Eve is trying to know about. If these states interact with a standard state $|u\rangle$,

$$|\psi\rangle|u\rangle \rightarrow |\psi\rangle|v\rangle$$
$$|\varphi\rangle|u\rangle \rightarrow |\varphi\rangle|v'\rangle$$

Eve would want $|v\rangle$ and $|v'\rangle$ to be different, to know the identity of the state. However inner products are preserved under unitary transformations and

$$\langle v|v'\rangle\langle \psi|\varphi\rangle = \langle u|u\rangle\langle \psi|\varphi\rangle \quad \text{or,} \quad \langle v|v'\rangle = \langle u|u\rangle = 1$$

So $|v\rangle$ and $|v'\rangle$ must be identical and Eve will need to disturb one of the two states in order to acquire any information.

## 4. Kak's three-stage quantum protocol

This protocol can be summarized as follows:

***Step 1:*** Alice applies a unitary transformation $U_A$ on quantum information X and sends the qubits to Bob.

***Step 2:*** Bob applies $U_B$ on the received qubits $U_A$, which gives $U_B U_A(X)$ and sends it back to Alice.

***Step 3:*** Alice applies $U_A^\dagger$ (transpose of the complex conjugate of $U_A$) on the received qubits to get $U_A^\dagger U_B U_A(X) = U_A^\dagger U_B U_A(X) = U_B(X)$ and sends it back to Bob.

Bob then applies $U_B^\dagger$ on $U_B(X)$ to get the information X.

Here $U_A$ and $U_B$ must be commutative to each other, which means that $U_B U_A(X) = U_A U_B(X)$.

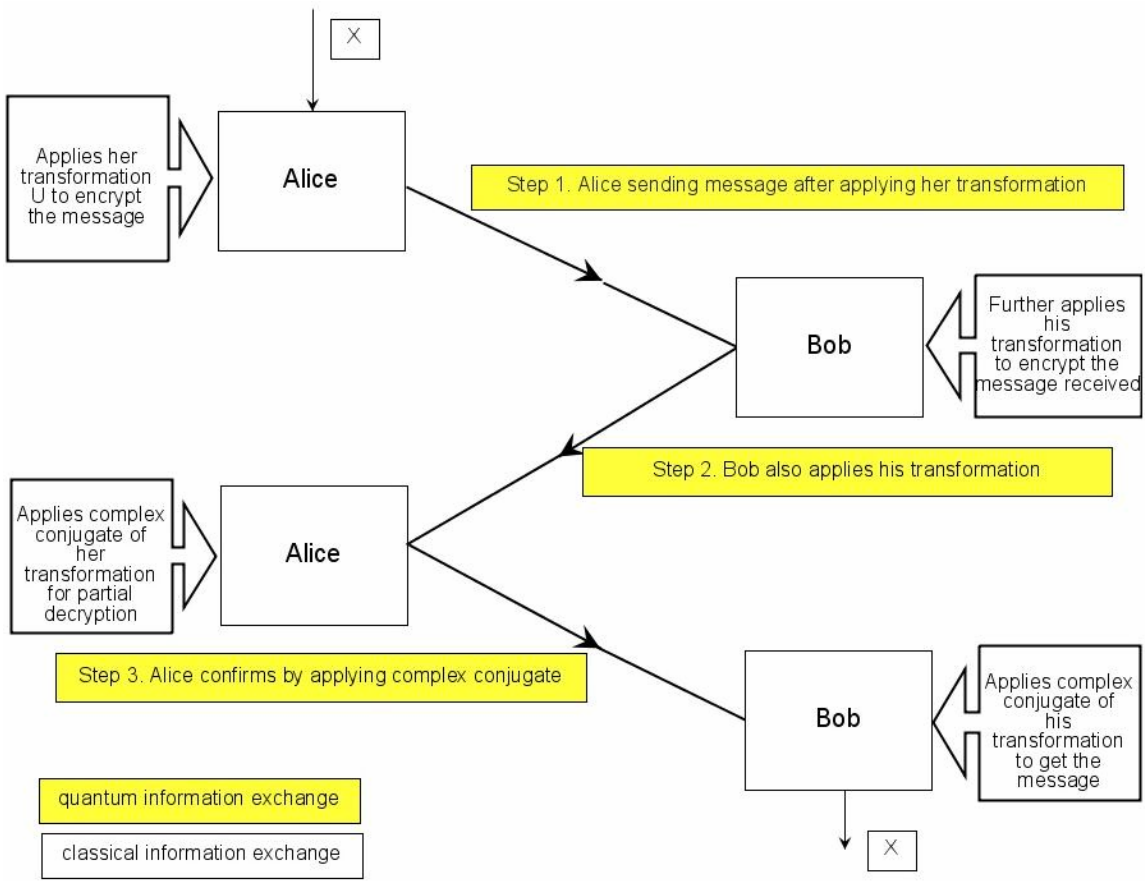

**Fig 2. Kak's three-stage protocol**

With n number of qubits present in the message, the transformations $U_A$ and $U_B$ both must be of $2^n$ dimension. It has been observed that the (2 × 2) rotation operator, Pauli-X, Pauli-Y and Pauli-Z can be used as commutative transformations in 1-qubit system as all of these are 2 × 2 matrices.

In order to find transformations for an n-qubits system we can randomly pick any of these 2×2 matrices and tensor multiply it [7] with another randomly picked one (may be itself) and keep on tensor multiplying for n times, which will eventually produce a $2^n \times 2^n$ matrix. The commutativity of the rotation operator can be shown as below:

$$R(\theta) = \begin{pmatrix} \cos\theta & -\sin\theta \\ \sin\theta & \cos\theta \end{pmatrix}$$

$$R(\theta).R(\phi) = \begin{pmatrix} \cos\theta & -\sin\theta \\ \sin\theta & \cos\theta \end{pmatrix} \cdot \begin{pmatrix} \cos\phi & -\sin\phi \\ \sin\phi & \cos\phi \end{pmatrix} = \begin{pmatrix} \cos(\theta+\phi) & -\sin(\theta+\phi) \\ \sin(\theta+\phi) & \cos(\theta+\phi) \end{pmatrix}$$

For a 2-qubit system:

$$R_2(\theta) = \begin{pmatrix} 1 & 0 \\ 0 & 1 \end{pmatrix} * \begin{pmatrix} \cos\theta & -\sin\theta \\ \sin\theta & \cos\theta \end{pmatrix} = \begin{pmatrix} \cos\theta & -\sin\theta & 0 & 0 \\ \sin\theta & \cos\theta & 0 & 0 \\ 0 & 0 & \cos\theta & -\sin\theta \\ 0 & 0 & \sin\theta & \cos\theta \end{pmatrix}$$

where "*" denotes tensor product.

Keeping the practical implementations in view information exchange in three-stage protocol does not restrict to single photon. Even if the laser pulse produces multiple photons and as long as all the photons are in the same phase the transformation and their complex conjugate transformation will have same effect on them. So irrespective of how many photons are used three-stage protocol is bound to succeed and is not prone to beam-splitting attack.

But, theoretically, the three-stage protocol can be subjected to the man-in-the-middle attack [5]. The next section explains how it can affect functionality of three-stage protocol.

### 5. Man-in-the-middle attack for three-stage protocol

Man-in-the-middle attack can affect both classical and quantum channels. Here Eve can pretend to be Bob to Alice and vice-versa. Instead of $U_B$ Eve selects $U_C$ (which is also commutative) and fakes a response which looks similar to what Bob would have done. Eve pretends as Alice to Bob with the transformation $U_D$, which is commutative to $U_B$ and instead of X sends a gibberish Y. So from interaction with Alice he acquires value X and sends a junk Y to Bob and hence disables the protocol.

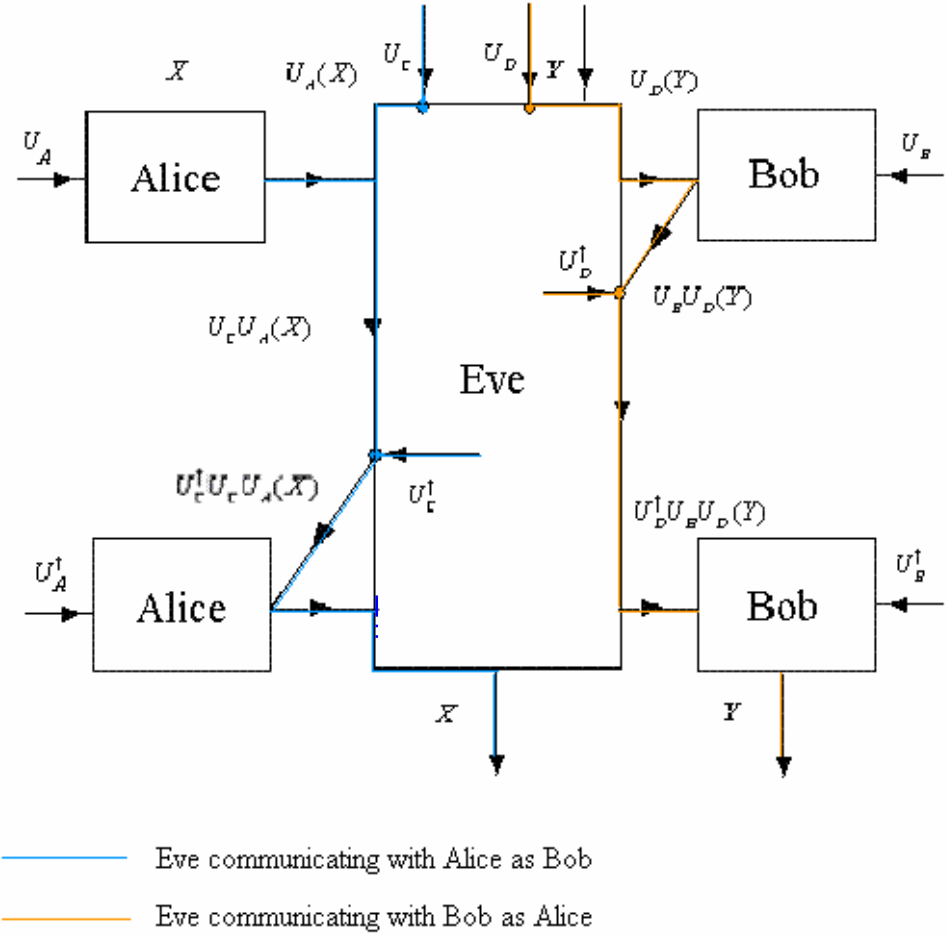

**Figure 3. Three-stage protocol under Man-in-the-middle attack**

EPR pairs can be used for a possibly secure three-stage protocol that can avoid man-in-the-middle attack [9]. But while distributing the EPR pairs, they might get corrupt during the transit. So if the EPRs are not matched then the process will be aborted and this might cause delay. So in order to ensure security from man-in-the-middle attack and avoid the uncertainty regarding EPR pairs this paper proposes a hybrid model that uses classical authentication protocols to ensure security in three-stage quantum protocol.

**6. Description of the classical authentication aided protocol**

Here, we use modified version of the protocols proposed by Denning-Sacco [9] and Kehne et al [10], as the authentication protocol, alongside the qubits sent in each stage. It takes help from a central Key Distribution Center (KDC), which assigns the session key and work as the central authority for authentication.

Firstly, as this protocol uses classical bit sequence, we have to transform the bit sequence into qubits. A bit sequence of 01101… can be transformed into $|0\rangle|1\rangle|1\rangle|0\rangle|1\rangle…$, even; to increase the amount of reliability we can map 0 and 1 into more than one photon. Now what we are doing in each step is that we are sending a series of photon as in usual three-stage protocol and still using the authentication protocol. Each time Alice or Bob (or the KDC) gets the stream of photons; they convert the authentication part to classical information, then process it and again transform it into quantum information before transmitting. Say, Q (.) is the function used to denote conversion of classical bits to quantum qubits. $Q^{-1}(.)$ is also used to get the classical bits inside the Alice, Bob and KDC units and are not shown in the protocol.

The protocol can be described as follows:

1. $A \rightarrow B:$    $Q(ID_A \| N_a) \| U_A(X)$

2. $B \rightarrow KDC:$ $Q(ID_B \| N_b \| E_{K_b}[ID_A \| N_a \| T_b]) \| U_B U_A(X)$

3. $KDC \rightarrow A:$ $Q(E_{K_a}[ID_B \| N_a \| K_s \| T_b]) \| E_{K_b}[ID_A \| K_s \| T_b]) \| N_b) \| U_B U_A(X)$

4. $A \rightarrow B:$    $Q(E_{K_b}[ID_A \| K_s \| T_b]) \| E_{K_s}[N_b]) \| U_A^\dagger U_A U_B(X)$

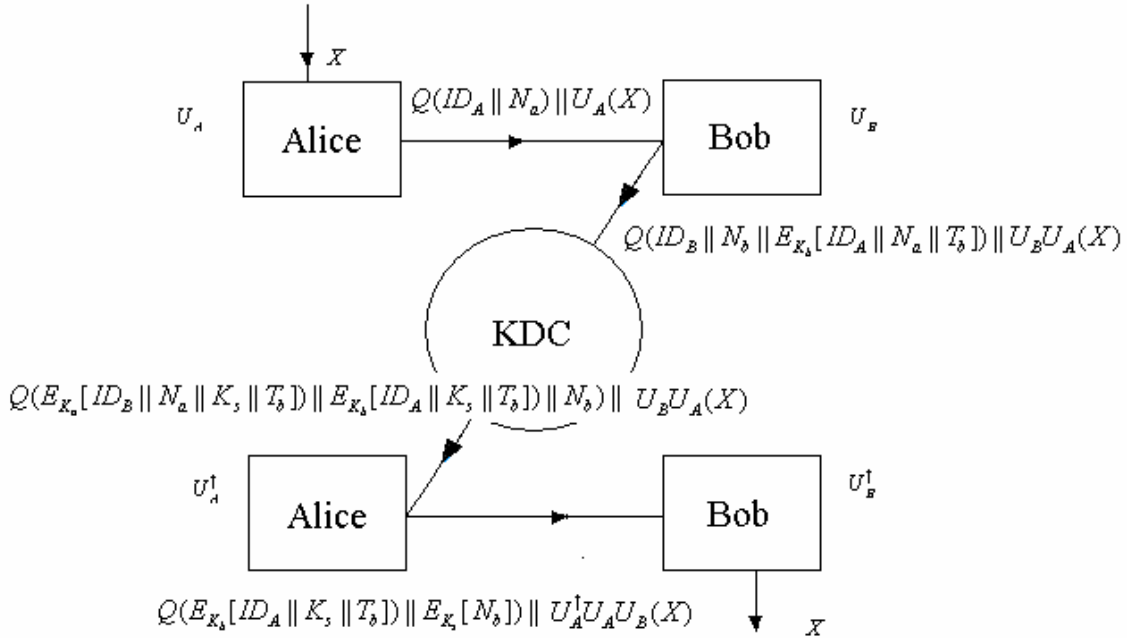

**Figure 4: Classical authentication protocol aided three-stage quantum protocol**

**1.** A sends quantum information of the nonce $N_a$ and the ID along side the message to B.

**2.** B also sends his ID and nonce and encrypts A's ID, nonce and his timestamp using shared key between B and KDC. This initiates KDC to assign a session key.

**3.** KDC assigns a session key and prepares packages for A and B which include their own ID, session key and B's timestamp. A's nonce is encrypted inside A's package but B's nonce is kept open. A gets back his nonce and A is assured of its timeliness by the session key and ensured that it's not a replay. This block also verifies that B has received A's earlier message with help of B's ID.

**4.** Session key authenticates that the message came from A and is not a replay.

## 7. Conclusion

The difficulty of generating single photons makes Kak's three-stage protocol more secure than other QKD protocols like BB84, which require use of single photons. Since all the photons, be it a single photon or multiple photons, go through private transformations in the Kak protocol, the information remains protected until the complex conjugate transformation is applied. In contrast, beam-splitting can easily break the inherent quantum security in BB84 and in order to avoid it measures may be taken that consume both time and money. Thus Kak's three-stage protocol may be used with greater confidence in its unbreakability than BB84.

Although Kak's protocol is secure against beam-splitting it can be successfully subjected to man-in-the-middle attack. To deal with such an attack timestamps, IDs, session keys, nonces and encryption keys are used for verification. If the authentication process detects any problem, the process is aborted. In this protocol we are dependent on the KDC and we need the KDC to be trustworthy.

There are many inherent advantages of the proposed system. The KDC can't see the message in this system. Neither Bob nor Eve is capable of modifying or forging the message. Alice cannot disavow sending that message and simultaneously Bob cannot deny receiving the message later on. The transmission remains quantum and hence non-reproducible. Synchronization throughout the network is not needed as the time-stamp is provided by Bob and hence will correspond to Bob's clock only. Suppress-reply attacks can be avoided because the nonces, the recipients will choose, are unpredictable to the sender.